\documentclass[prd, amsfonts, twocolumn, nofootinbib, showpacs]{revtex4}
\usepackage{graphicx, epsfig}
\usepackage{color}
\usepackage{amsmath}
\usepackage{amssymb}
\newcommand{\be}{\begin{equation}}
\newcommand{\ee}{\end{equation}}
\newcommand{\bea}{\begin{eqnarray}}
\newcommand{\eea}{\end{eqnarray}}

\newcommand{\gapp}{\mathrel{\raise.3ex\hbox{$>$}\mkern-14mu
\lower0.6ex\hbox{$\sim$}}}
\newcommand{\lapp}{\mathrel{\raise.3ex\hbox{$<$}\mkern-14mu
\lower0.6ex\hbox{$\sim$}}}
\def\bbox{{\,\lower0.9pt\vbox{\hrule \hbox{\vrule height 0.2 cm
\hskip 0.2 cm \vrule  height 0.2 cm}\hrule}\,}}

\begin{document}
\title{Can $\Lambda$CDM model reproduce MOND-like behavior?}
\author{De-Chang Dai, Chunyu Lu}
\affiliation{ Institute of Natural Sciences, Shanghai Key Lab for Particle Physics and Cosmology, School of Physics and Astronomy, Shanghai Jiao Tong University, Shanghai 200240, China}
 %%%%%%%%%%%%%%%%%%%%%%%%%%%%%%%%%%%%%%%%%%%%%%%%%%%%%%%

\begin{abstract}
\widetext

It is usually believed that MOND can describe the galactic rotational curves with only baryonic matter and without any dark matter very well, while the $\Lambda$CDM model is expected to have difficulty in reproducing MOND-like behavior. Here, we use EAGLE's data to learn whether the $\Lambda$CDM model can reproduce MOND-like behavior. EAGLE's simulation result clearly reproduces the MOND-like behavior for $a_b\gtrapprox 10^{-12}\text{m/s}^2$ at $z=0$, although the acceleration constant, $a_0$, is a little larger than the observational data indicate. We find that $a_0$ increases with the redshift in a way different from what Milgrom proposed ($a_0\propto H$). Therefore, while galaxy rotation curves can be fitted by MOND's empirical function in the $\Lambda$CDM model, there is no clear connection between $a_0$ and the Hubble constant. We also find that $a_0$ at $z\gtrapprox 1$ is well separated from $a_0$ at $z=0$. Once we have enough galaxies observed at high redshifts, we will be able to rule out the modified gravity model based on MOND-like empirical function with a z-independent $a_0$.

\end{abstract}

%%%%%%%%%%%%%%%%%%%%%%%%%%%%%%%%%%%%%%%%%%%%%%%%%%

\maketitle

\section{Introduction} \label{sec:intro}

Modified Newtonian Dynamics(MOND) was proposed by Milgrom \citep{1983ApJ...270..365M} to explain the discrepancy between Newtonian dynamical mass and the directly observable mass in the galaxy scale. The best known example are the galactic rotation curves, which tend to become roughly flat ($V\sim$ constant) in the region where they are predicted to be Keplerian ($V\propto r^{-1/2}$). Today  this phenomenon is considered to be an evidence of the existence of dark matter. This explanation however also raises some doubts, because dark matter has never been directly observed. In addition, it appears that $\Lambda$CDM model must be fine-tuned to reproduce the behavior of the observed galactic rotation curves \citep{2015CaJPh..93..250M}. Therefore, though MOND is just an empirical theory, it still motivates several different modified gravity theories, e.g. TeVeS \citep{2004PhRvD..70h3509B}.

Basically, MOND introduces a new constant $a_0$ which defines a boundary between the strong and the weak gravity region (or deep-MOND region). The dynamics in the strong gravity region is  essentially the same as the Newtonian gravity, while the dynamics  in the weak gravity region must be modified. The constant rotational velocity in a galaxy's outskirts is caused by this modified dynamics. Although a constant $a_0$ explains the galaxies' rotation curves very well, Milgrom also noticed that $a_0$ is approximately equal to $cH_0/2\pi$ ($H_0$ is today's Hubble constant) and proposed that $a_0$ may change with the local Hubble constant\citep{1983ApJ...270..365M}.

Recently there had been several studies on galaxy kinematics at high redshifts\citep{2016ApJ...819...80P,2016ApJ...831..149W,2017ApJ...840...92L}. It has been found that there was very little dark matter in the high redshift galaxies\citep{2017Natur.543..397G}. This could be a fatal failure for MOND. However, Milgrom found that those studies are still in the strong gravity region\citep{2017arXiv170306110M} and cannot rule out MOND. However, they can provide a constraint on $a_0$ at high redshifts.  He found that $a_0$ at $z=2$ must be less than four times of the value of $a_0$ at $z=0$. Since more and more galaxies have been found at higher redshifts, the data will be able to distinguish between $a_0$ as a universal constant and a redshift dependent parameter. Therefore it is warranted  to test whether $\Lambda$CDM model reproduces MOND-like behavior, and to study how $a_0$ changes with redshift in the $\Lambda$CDM model. To do this, we use EAGLE's simulation dataset\citep{2015MNRAS.446..521S,2015MNRAS.450.1937C,2016A&C....15...72M}. Although it is widely believed that $\Lambda$CDM needs a fine-tuned initial condition to reproduce MOND-like rotation curves, we find that the EAGLE's galaxies' dynamics is very similar to what MOND predicts for $a_b \gtrapprox 10^{-12}\text{m/s}^2$. The simulation already provides  a quite reasonable result even with the present data. Therefore, the claim that $\Lambda$CDM needs to be fine-tuned to reproduce MOND behavior might not be correct. Our study also indicates that $a_0$ is increasing with the redshift, and apparently cannot be considered to be a fluctuation at $z\gtrapprox 1$. If this result is confirmed by  observations, then some of the modified gravity models based on a constant $a_0$ will be ruled out, e.g. TeVeS \citep{2004PhRvD..70h3509B}..

In the following we introduce the EAGLE dataset and the basic idea of MOND. We then fit the total kinetic acceleration with the MOND's empirical equation. At the end we show how many galaxies are needed to achieve the necessary statistical significance in order to rule out $a_0$ as a z-independent constant.

\section{Galaxy simulation}

There are several public cosmological hydrodynamical simulations\citep{2005MNRAS.364.1105S,2015A&C....13...12N,2016A&C....15...72M}. We use EAGLE simulation suite\citep{2015MNRAS.446..521S,2015MNRAS.450.1937C,2016A&C....15...72M}. EAGLE simulation starts from a redshift of $z=127$ to present day. The simulation adopts a flat $\Lambda$CDM cosmology with parameters taken from the Planck result: $\Omega_\Lambda =0.693$, $\Omega_m=0.307$, $\Omega_b=0.04825$, $\sigma_8=0.8288$, $n_s=0.9611$, $Y=0.248$ and $H_0=67.77\text{km/s}^2$. The mass component includes gas, stars, black holes and dark matter. EAGLE provides 6 simulations with 29 different snapshots. We show only two simulation datasets, RefL0025N0752 and RefL0100N1504, in this article.  RefL0025N0752 has the finest particle mass resolution and  RefL0100N1504 has the largest box's size. The other four datasets are similar to these two, and we do not show them. The details of the simulation parameters and other technical issues can be found in EAGLE's public web site and \citep{2016A&C....15...72M}.

\begin{figure}
   \centering
\includegraphics[width=10cm]{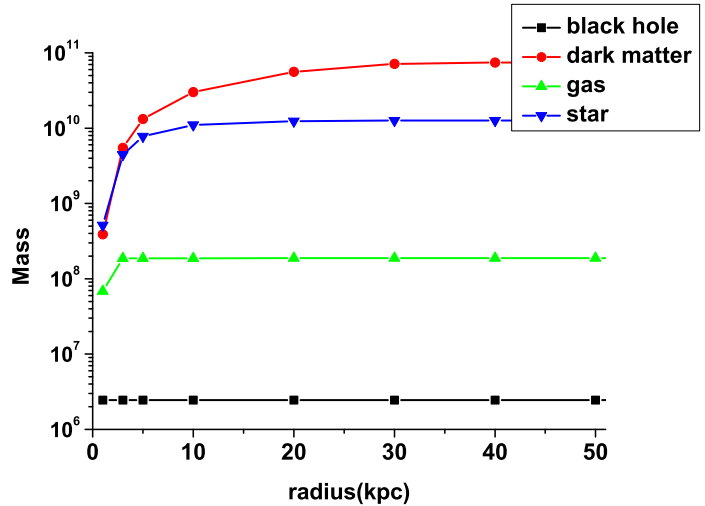}
\caption{The mass distribution in a galaxy in RefL0025N0752 dataset. The redshift is $z=0$. There is more regular matter (stars, gas and black holes) than dark matter at short radius. There is more dark matter than regular matter at larger radius.
}
\label{Mass}
\end{figure}

Figure \ref{Mass} shows a typical galaxy's mass distribution. There is more regular matter near the galaxy's center. This is the strong gravity region in MOND ($a>a_0$) where the gravitational acceleration is the same as the Newtonian gravity. There is more dark matter farther away from the galaxy's center. This is the weak gravity region where the gravitational acceleration is different from the Newtonian gravity and MOND can play some role. The mass is increasing with the radius at short distances and becomes almost a constant at larger distances, at which MOND also fails to explain gravitational acceleration with just baryonic matter and must be avoided in this study. We use galaxies with stellar mass, $M_*$, between $5\times 10^{9}M_\odot$ and $5\times 10^{10}M_\odot$, because EAGLE's simulation shows that the galaxy rotation curve has clear discrepancy from observation data for galaxies with $M_*>10^{11}M\odot$, and particle mass resolution may affect lighter galaxies', $M_*<10^9 M_\odot$, behavior \citep{2015MNRAS.451.1247S}. The galaxies' total dynamical acceleration are obtained from

\begin{equation}
a_t=\frac{G\Big(M_g(r)+M_*(r)+M_{bh}(r)+M_d(r)\Big)}{r^2} .
\end{equation}
Here $G$ is the Newton's constant, $M_g(r)$, $M_*(r)$, $M_{bh}(r)$ and $M_d(r)$ are gas mass, stellar mass, black hole mass and dark matter mass within radius $r$ respectively. In MOND the central acceleration can be determined by the baryon matter's acceleration,

\begin{equation}
a_b=\frac{G\Big(M_g(r)+M_*(r)+M_{bh}(r)\Big)}{r^2} .
\end{equation}

\begin{figure}
   \centering
\includegraphics[width=10cm]{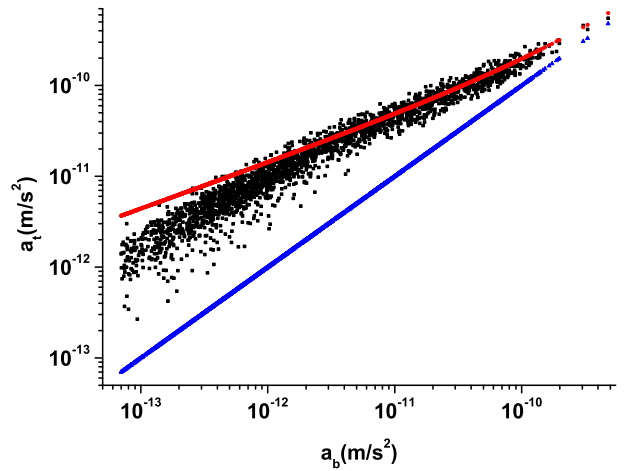}
\caption{The black dots represent the total acceleration, $a_t$, with respect to the baryon's acceleration ($a_b$). We choose galaxies with baryon mass between $5\times 10^{10}M_\odot$ and $5\times 10^{11}M_\odot$. The data is from RefL0025N0752, and redshift is $z=0$. The red curve is plotted according to the function $a=(a_b+\sqrt{a_b^2+4\times 1.9\times 10^{-10}a_b})/2$. The black dots change with $a_b$ in the same way as the red curve for $a_b \gtrapprox 2\times 10^{-12}$. The black dots decrease much faster than the red curve for $a_b \lessapprox 2\times 10^{-12}$. MOND cannot provide an appropriate empirical function in this region, so we avoid it. The blue line is for $a=a_b$. The black dots are above the blue line, because of the existence of dark matter.
}
\label{acceleration}
\end{figure}

Figure \ref{acceleration} shows the relation between $a_t$ and $a_b$. It is clear that $a_t$ is always larger than $a_b$, because of the existence of dark matter. The discrepancy can be clearly seen from $a_b\lessapprox 10^{-9}\text{m/s}^2$. $a_t$ and $a_b$ follow the MOND-like behavior for $a_b\gtrapprox 10^{-12}\text{m/s}^2$ and then is discrepant from MOND-like for $a_b\lessapprox 10^{-12}\text{m/s}^2$, because dark matter's mass does not increase as quickly as it does at shorter distances. Therefore we must not include the low $a_b$ region, because MOND cannot properly describe gravitational acceleration in this region. Most observations do not have such low acceleration data anyway.

\section{MOND}

The basic idea of MOND is to explain how the galaxy rotation curve becomes flat in the weak gravity region (for $a<a_0$). To achieve a constant rotational velocity in galactic outskirts, Milgram proposed that the true gravitational acceleration, $a$, is related to the Newtonian gravitational acceleration, $a_b$, in the form \citep{1983ApJ...270..365M}

\begin{equation}
\mu(a/a_0)\vec{a}=\vec{a}_b .
\end{equation}

Here $a_0$ is a critical acceleration constant, and $\mu$ is the MOND interpolating function, which satisfies

\begin{eqnarray}
\mu(x) =\left\{
\begin{array}{cc}
1 & \text{if } a\gg a_0,
\\
\frac{a}{a_0} & \text{if } a\ll a_0 .
\end{array}\right.
\end{eqnarray}

%\begin{eqnarray}
%\gamma^\mu  & = &
% \Bigg(
%\begin{array}{cc}
%0 & \sigma^\mu_+ \\
%\sigma^\mu_- & 0
%\end{array}
%\end{eqnarray}

There are several different versions of $\mu$ in the literature. Here we choose
\begin{equation}
\mu (x)=\frac{x}{1+x} .
\end{equation}

One then finds dynamical gravitational acceleration in the MOND from the baryonic matter.
\begin{equation}
\label{MOND1}
a=\frac{a_b+\sqrt{a_b^2+4 a_b a_0}}{2} .
\end{equation}

Although this empirical function successfully reproduces galaxies' rotation curves, it is considered to be only a successful phenomenological scheme, instead of a theory. Bekenstein and Milgrom realized that a Lagrangian based theory of MOND can be achieved by introducing a scalar field\citep{1984ApJ...286....7B}. To calculate a galaxy's dynamical acceleration in the theory, one must solve the scalar field's equation of motion. However, Milgrom had shown that original MOND agrees with the scalar field equation within a few percent of discrepancy\citep{1986ApJ...302..617M}. Therefore, we are going to use the calculation from the original MOND to avoid complicated calculation.

Although Milgrom proposed a constant acceleration parameter $a_0$, in \citep{1983ApJ...270..365M,1989ComAp..13..215M} he also noticed that

\begin{equation}
a_0\approx \frac{cH_0}{2\pi} ,
\end{equation}

where $H_0$ is the present Hubble constant. This implies that $a_0$ may be originally connected to the Hubble constant, $H(z)$. In this case $a_0$ is a z-dependent function,

\begin{equation}
\label{H-dep}
a_0(z)=a_0(0)\sqrt{(\Omega_m+\Omega_b)(1+z)^3+\Omega_\Lambda} .
\end{equation}

Here the local Hubble constant, $H$, is obtained from the Friedmann equation, and the radiation component is ignored.  We will compare this equation with $a_0$ from EAGLE's data at different redshifts.

\begin{figure}
   \centering
\includegraphics[width=10cm]{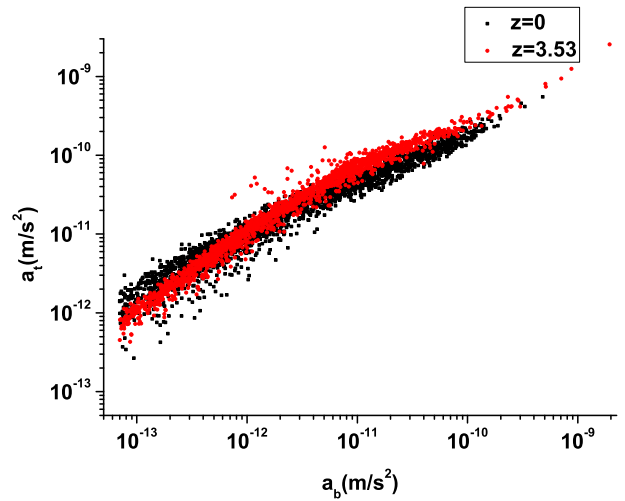}
\caption{The black dots are the total acceleration at z=0 as a function of $a_b$, and the red dots is the total acceleration at $z=3.53$ as a function of $a_b$. Only galaxies with baryonic mass in the region, $5\times 10^{10}M_\odot<M_b<5\times 10^{11}M_\odot$, in Ref-L0025N0752 are included. The red dots are above the black dots for larger $a_b$. Then the function decreases very quickly for smaller $a_b$. Apparently, the red dots do not match the black dots, and a single $a_0$ MOND-like empirical function is not enough to describe both the high redshift and low redshift acceleration behavior.
}
\label{compare-acceleration}
\end{figure}

Figure \ref{compare-acceleration} shows $a_t$ at different redshift. One finds $a_t$ is higher for higher redshift case. This implies $a_0$ also increases with $z$. Both $a_t$ at higher redshift and at lower redshift deviate from MOND in the very weak gravity region. In this article we consider only $a_b>10^{-11}\text{m/s}^2$ data points to avoid including the regions of deviation.

\section {Analysis}

\begin{figure}
   \centering
\includegraphics[width=10cm]{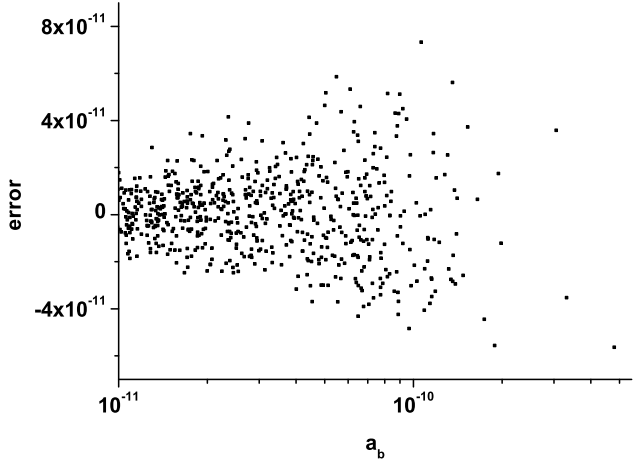}
\caption{This figure shows $a_t-a$ from Ref-L0025N0752 dataset. The redshift is $z=0$.
}
\label{error}
\end{figure}

\begin{figure}
   \centering
\includegraphics[width=10cm]{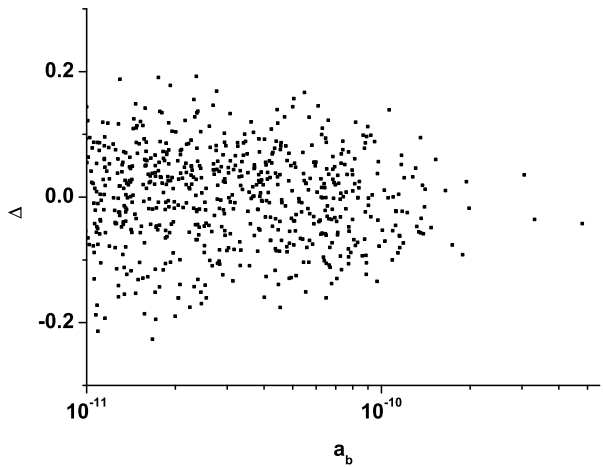}
\caption{This figure shows $\log_{10}a_t-\log_{10}a$ from RefL0025N0752 dataset. The redshift is $z=0$.
}
\label{error-log}
\end{figure}

We fit the total acceleration $a_t$ with equation \ref{MOND1} to find out the best fit for $a_0$. Since the simulations have no error bars, we assume that $a_t$'s standard deviation is the same for different $a_b$. However, if one compares $a_t$ with the best fit $a$, one finds that the fluctuation of $a_t-a$ is clearly $a_b$-dependent (figure \ref{error}), and that the fluctuation of $\log_{10}a_t-\log_{10}a$ is much more uniformly distributed (figure \ref{error-log}). Therefore,  $\chi^2$ is taken in the form

\begin{equation}
\chi^2(a_0)=\sum_i (\log_{10}a(i)-\log_{10}a_t(i))^2 ,
\end{equation}

where $i$ is the index for different data point. We  consider only $a_b >10^{-11}\text{m/s}^2$.

\begin{figure}
   \centering
\includegraphics[width=10cm]{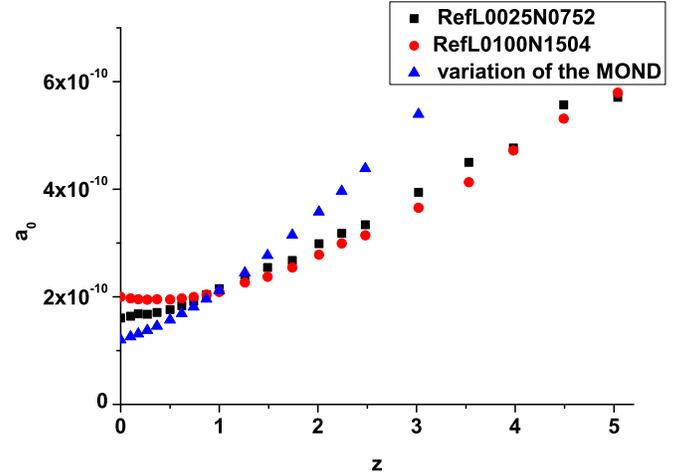}
\caption{This plot depicts how $a_0$ changes with respect to the redshift. $a_0$ is smaller at lower redshift and is larger at higher redshift. The black dots are from RefL0025N0752 and the red dots are from RefL0100N1504. The blue dots are from $a_0 =1.2\times 10^{-10} \sqrt{\Omega_m (1+z)^3+\Omega_\Lambda }$. Here, $\Omega_m =0.3$, $\Omega_\Lambda=0.7$. Apparently, blue dots are very different from the simulation result.
}
\label{a0-change}
\end{figure}

We minimize $\chi^2$ and find the best fit for $a_0$.  Figure \ref{a0-change} shows how $a_0$ changes with the redshift.  Both RefL0025N0752 and  RefL0100N1504 datasets show that $a_0$ increases with $z$. This result is consistent with recent high redshift galaxies observations \citep{2017arXiv170306110M}. However, for the redshifts between $z=0$ and $z=1$,   $a_0$ from RefL0100N1504 is almost a constant, but $a_0$ from  RefL0025N0752 is still increasing with $z$.  Whether this discrepancy at lower redshift is caused by the particle mass resolution or the other  model dependent parameters (like $\Delta T_{AGN}$) is unclear. Apart from this discrepancy, both simulations have a larger $a_0$ at $z=0$ than $a_0$ from observations ($\approx 1.2\times 10^{-10}\text{m/s}^2$).  Since RefL0025N0752's result is much closer to the real observations, we are going to focus on  RefL0025N0752 case. The other four EAGLE's simulation data give similar result like either RefL0025N0752 or RefL0100N1504, so we do not include them in the figure. One also finds that the assumption that $a_0$ is proportional to $H$( equation \ref{H-dep}) does not match the simulation very well. This implies that $a_0\propto H$ is not consistent with the $\Lambda$CDM model. One can also estimate the standard deviation from $\chi^2/N$, where $N$ is the number of the total data points. The standard deviation is  $0.06 \text{dex}\lessapprox \sigma\lessapprox  0.08\text{dex}$ for redshift from $0$ to $4$. This is about one half of the value from the real observation \citep{2016PhRvL.117t1101M}. This is not surprising, since there is always more sources of uncertainty in the real data. This also means that $\Lambda$CDM can reproduce MOND-like behavior. There is no miracle.

\begin{figure}
   \centering
\includegraphics[width=10cm]{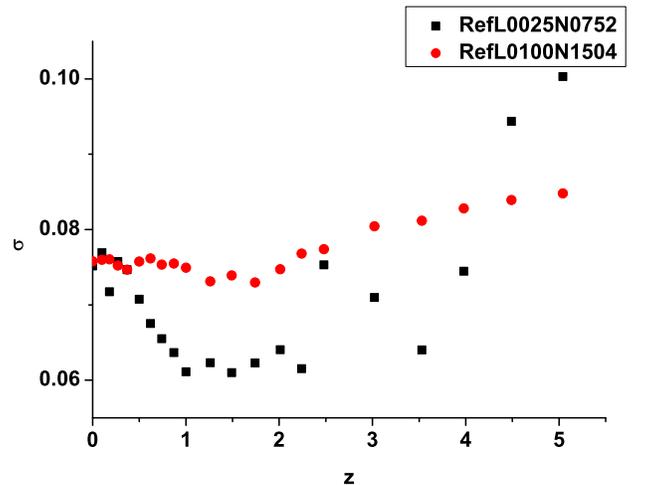}
\caption{  The standard deviation of the $a_t$ from RefL0025N0752 and RefL0100N1504 datasets.
}
\label{sigma-change}
\end{figure}

\begin{figure}
   \centering
\includegraphics[width=10cm]{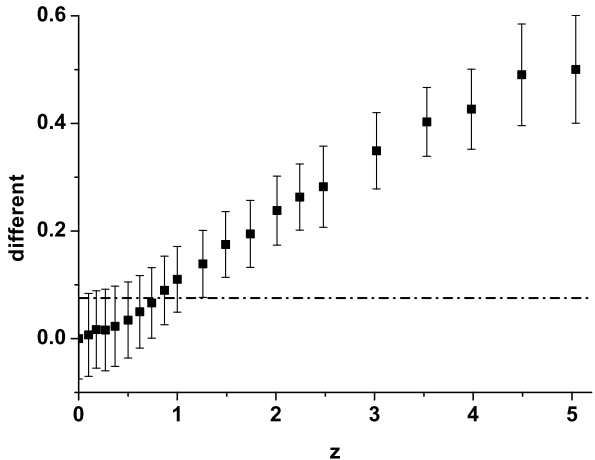}
\caption{The black squares show $\log_{10}a(z,a_b=10^{-11})-\log_{10}a(z,a_b=10^{-11})$. The vertical line is the standard deviation (not the error bar) of the total acceleration. The result is from the simulation RefL0025N0752.
}
\label{rule}
\end{figure}

\begin{figure}
   \centering
\includegraphics[width=10cm]{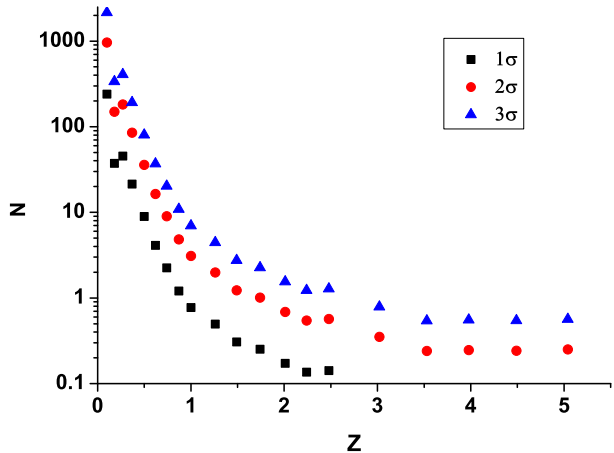}
\caption{The black, red,  and blue dots represent the number of galaxies needed in order to rule out a z-independent MOND-like empirical function $a_0$  at $1 \sigma$, $2 \sigma$ and $3\sigma$ significance. The result is from the fitting parameter of simulation RefL0025N0752 and $a_b$ is taken to be $10^{-11}\text{m/s}^2$.
}
\label{number}
\end{figure}

Now, we want to estimate under what conditions the MOND's acceleration constant can not be considered to be a redshift independent parameter. We take $a_b=10^{-11}\text{m/s}^2$ and plot the expected accelerations with their standard deviations, $\sigma$, at different redshift (figure \ref{rule}). This is in the deep-MOND region and is similar to McGaugh's method\citep{2011PhRvL.106l1303M}, which has standard deviation about 4 times as big as this analysis. While $z>1 $, the expected accelerations are well separated and can be easily distinguished from each other. One can also estimate how much data is needed to distinguish the $a_0(z)$ from $a_0(0)$ from the Z-test.

$Z=\frac{\log_{10}a(z,a_b=10^{-11})-\log_{10}a(z=0,a_b=10^{-11})}{\sqrt{\frac{\sigma(z)^2}{n}+\frac{\sigma(z=0)^2}{m}}}$.

Here $n$ is the number of data points at the redshift $z$, and $m$ is the number of data points at $z=0$. We take $n=m$ just for convenience. In the real case, $m$ can be much bigger than $n$, because there are much more galaxies found at the lower redshifts. Figure \ref{number} shows that less than 10 data points are needed to distinguish $a_0(z=1)$ from $a_0(z=0)$ at $3\sigma$ significance. Since there are more data points for one single galaxy in a real situation, the number of the needed galaxies can be lower. However, the standard deviation in a real observation is larger than what we have here, so the galaxy number which is actually needed is unclear, but should be about the same order.

\section{Conclusion}

It is usually believed that the standard $\Lambda$CDM model cannot explain why the galactic rotation curves are MOND-like\citep{2015CaJPh..93..250M}. However, we show that the dynamics of galaxies in EAGLE's simulation is well described by MOND alone for some values of gravitational acceleration. This indicates that $\Lambda$CDM has no clear conflict with MOND (at least for $a_b\gtrapprox 10^{-12}\text{m/s}^2$ at $z=0$) and general belief that $\Lambda$CDM cannot produce MOND behavior and a modified gravity model is needed might not be correct. We also find that $a_0$ increases with the redshift (figure \ref{a0-change}). This relation is not the same as what Milgrom proposed $a_0\propto H$\citep{1983ApJ...270..365M}. EAGLE's simulation shows that for $z\gtrapprox 1$, $a_0$ increases approximately linearly with $z$ for $z>1$. However, the simulations do not agree with each other for $z\lessapprox 1$ and they give larger values for $a_0$ than the observations at $z=0$ does.

Since the result from RefL0025N0752 is much closer to  observations at $z=0$, we focus on this simulation and compare how $a_0$ changes with the redshift. One can see that $a_0 (z\gtrapprox 1)$ is well separated from $a_0(z=0)$. This variation is well within the result from the recent high redshift galaxies \citep{2017arXiv170306110M}. $3\sigma$ significance can be achieved  with less than 10 data points for $a_b=10^{-11}\text{m/s}^2$ at $z=1$. This is achievable since there are much more higher redshift galaxies found recently. However, it is unclear how many galaxies are needed to clarify the redshift dependent trends since the actual data could have much larger fluctuations, and also one single galaxy can provide more data points.  But we expect it is of the same order of magnitude as in our analysis.

\begin{acknowledgments}
D.C Dai was supported by the National Science Foundation of China (Grant No. 11433001 and 11447601), National Basic Research Program of China (973 Program 2015CB857001), the key laboratory grant from the Office of Science and Technology in Shanghai Municipal Government (No. 11DZ2260700) and  the Program of Shanghai Academic/Technology Research Leader under Grant No. 16XD1401600. We acknowledge the Virgo Consortium for making their simulation data available. The eagle simulations were performed using the DiRAC-2 facility at Durham, managed by the ICC, and the PRACE facility Curie based in France at TGCC, CEA, Bruyeresle-Chatel
\end{acknowledgments}

\end{document}